\documentclass[10pt]{article}
\usepackage{multicol}
\usepackage{graphicx}
\usepackage{amsmath}
\usepackage[a4paper]{geometry}
\usepackage{hyperref}
\usepackage{rotating}

\renewcommand{\baselinestretch}{1.15}

\begin{document}

\begin{center}
{\Large \bf Analyzing transverse momentum spectra by a new method in
	high-energy collisions}

\vskip.75cm

Li-Li~Li$^1${\footnote{E-mail: shanxi-lll@qq.com;shanxi-lll@sxau.edu.cn}}, Fu-Hu~Liu$^{2,}${\footnote{E-mail: fuhuliu@163.com;
fuhuliu@sxu.edu.cn}}, Muhammad~Waqas$^3$ and Muhammad Ajaz$^{4}$

\vskip.25cm

{\small\it $^1$Department of Basic Sciences, Shanxi Agricultural
	University, Jinzhong 030801, China\\

$^2$Institute of Theoretical Physics, State Key Laboratory of Quantum Optics and Quantum Optics Devices \&
Collaborative Innovation Center of Extreme Optics, Shanxi
University, Taiyuan~030006, China\\

$^3$ School of Nuclear Science and Technology, University
of Chinese Academy of Sciences, Beijing 100049, China\\

$^4$ Department of Physics, Abdul Wali Khan University
Mardan, Mardan 23200, Pakistan}

\end{center}

\vskip.5cm

{\bf Abstract:} {We analyzed the transverse momentum spectra of
	positively and negatively charged pions ($\pi^+$ and $\pi^-$),
	positively and negatively charged kaons ($K^+$ and $K^-$), protons
	and antiprotons ($p$ and $\bar p$), as well as $\phi$ produced in
	mid-(pseudo)rapidity region in central nucleus--nucleus (AA)
	collisions over a center-of-mass energy range from 2.16 to 2760
	GeV per nucleon pair. The transverse momentum of the considered
	particle is regarded as the joint contribution of two participant
	partons which obey the modified Tsallis-like transverse momentum
	distribution and have random azimuths in superposition. The
	calculation of transverse momentum distribution of particles is
	performed by the Monte Carlo method and compared with the
	experimental data measured by international collaborations. The
	excitation functions of effective temperature and other parameters
	are obtained in the considered energy range. With the increase of
	collision energy, the effective temperature parameter increases
	quickly and then slowly. The boundary appears at around 5 GeV,
	which means the change of reaction mechanism and/or generated
	matter.
\\

\\


\vskip1.0cm

{\section{Introduction}}

The space-time evolution of hadron--hadron, hadron--nucleus, and
nucleus--nucleus (AA) or heavy-ion collisions is a complex process
which involves different degrees of freedom under different
spatiotemporal coordinates. Because of this complexity, it is
difficult to use a theory to describe the development of the
entire system. After the initial stage of heavy-ion collisions,
the system undergoes to a pre-equilibrium phase, followed by the
de-confined quark-gluon plasma (QGP) phase and then a possible
mixing phase, in which it should display at least a signal of the
first-order phase transition. The hadronization then takes place
where the compound hadrons are formed from the original partons.
With the increase of collision energy, the energy or temperature
at which the phase transition from hadron to QGP may occur
initially is referred to as the critical energy or temperature.

After the hadronization stage, the chemical composition of the
system is frozen and inelastic collisions stop, where the particle
ratios are fixed. Immediately afterwards, with the expansion of
the system, the mean-free-path of the particles becomes larger
than the size of the system, and this stage is referred as the
kinetic freeze-out stage. The transverse momentum ($p_T$) spectra
of particles are no longer changed. Finally, the particles fly to
the detector and their properties are measured. The temperature at
the stage of chemical freeze-out is called the chemical freeze-out
temperature ($T_{ch}$)~\cite{a,b,c}, and the stage of
kinetic freeze-out is known as the kinetic freeze-out temperature
($T_0$ or $T_{kin}$).

We are interested in the study of particles at the stage of
kinetic freeze-out. The kinetic freeze-out is an important and
complex issue. Different literature presented different kinetic
freeze-out scenarios such as the single~\cite{d},
double~\cite{e,f}, triple~\cite{g}, and multiple kinetic freeze-out
scenarios~\cite{h,i,j}. In addition, the behavior of $T_0$ with
increasing the centrality and collision energy is also very
complex~\cite{k,l,m,n}. To our knowledge, the behavior of $T_0$
with the collision energy is known to increase from a few GeV to 7
or 10 GeV, after which the trend becomes indefinitely saturated,
increscent, or decrescent. This indefinite trend is caused by the
different exclusions of flow effect. Different from $T_0$, the
effective temperature $T$ which contains the contributions of
thermal motion and flow effect has definite behavior. Therefore,
we focus our attention on the energy dependence of $T$ in central
AA collisions.

In this work, we will use a new method to analyze the $p_T$
spectra of particles so that we can extract $T$ and other
parameters. The value of $p_T$ for a given particle can be seen as
the superposition of contributions of two participant partons with
random azimuths, where the two partons are from the projectile and
target nuclei generally. In the rest frame of the emission source,
partons are assumed to emit isotropically. The Monte Carlo method
is performed and the statistical treatment is used in the fit to
the spectra. The transverse momentum contributed by each parton is
assumed to obey the modified Tsallis-like distribution. Thus, the
particle's $p_T$ is obtained from the synthesis of two vectors
with different sizes and~directions.

In order to verify our results, the $p_T$ (or the transverse mass
$m_T$) spectra of positively and negatively charged pions ($\pi^+$
and $\pi^-$), positively and negatively charged kaons ($K^+$ and
$K^-$), protons and antiprotons ($p$ and $\bar p$), as well as
$\phi$ produced at mid-(pseudo)rapidity (mid-$y$ or mid-$\eta$)
measured in central gold-gold (Au-Au) collisions at the heavy-ion
accelerator SIS (Schwerionensynchrotron) at the GSI (Gesellschaft
f{\"u}r Schwerionenforschung) in Darmstadt, Germany, by the
KaoS~\cite{22a} and HADES~\cite{22b,22c} Collaborations, in
central Au-Au collisions at the Alternating Gradient Synchrotron
(AGS) at BNL (Brookhaven National Laboratory) in Upton, USA, by
the E866~\cite{23}, E895~\cite{24,25}, and E802~\cite{26,27}
Collaborations, in central Au-Au collisions at the Relativistic
Heavy Ion Collider (RHIC) at BNL by the STAR~\cite{28,29,30,37,38}
and PHENIX~\cite{31,32} Collaborations, as well as in central
lead--lead (Pb-Pb) collisions at the Large Hadron Collider (LHC) at
CERN (Conseil Europ{\'e}enn pour la Recherche Nucl{\'e}aire) in
Geneva, Switzerland, by the ALICE Collaboration~\cite{36,21,39}
are studied. We can fit the data and extract the excitation
functions (energy dependences) of~parameters.

The remainder of this paper is structured as follows: the
formalism and method are shortly described in Section 2. Results
and discussion are given in Section 3. In Section 4, we summarize
our main observations and conclusions.
\\

{\section{Formalism and Method}}

According to Refs.~\cite{1,2,3}, one has the joint density
function of $y$ and $p_T$ in terms of the Tsallis-like
distribution at mid-$y$ to be
\begin{align}
	\frac{d^2N}{dydp_T} \propto \frac{dN}{dy} m_T
	\bigg[1+\frac{(q-1)(m_T-\mu-m_0)}{T}\bigg]^{-q/(q-1)},
\end{align}
where $N$ is the number of particles, $m_T =\sqrt{p_T^2+m_0^2}$,
$m_0$ is the rest mass of a given particle, $q$ is the entropy
index that characterizes the degree of equilibrium or
non-equilibrium, and $\mu$ is the chemical potential. Generally,
the mentioned joint density function is normalized to $N$. If
needed, the joint density function can be transformed to the
probability density functions of $y$ and $p_T$, respectively, which
are normalized to 1, respectively. It should be noted that Equation (1)
is an experiential expression obtained by us, in which $m_T$ in
front of the bracket replaced $p_T$ in the Tsallis distribution
due to our attempts. Equation (1) is not an ad-hoc version of the
Tsallis distribution used in literature~\cite{1,2,3}, though it is
very similar to the later. Thus, we call it the Tsallis-like
distribution which is suitable in the following~calculations.

According to our recent work~\cite{4}, by fitting the $p_T$
spectrum, in terms of the probability density function $f(p_T,T)$,
Equation (1) can be revised as
\begin{align}
	f(p_T,T)=\frac{1}{N}\frac{dN}{dp_T} = C m_T^{a_0}
	\bigg[1+\frac{(q-1)(m_T-\mu-m_0)}{T}\bigg]^{-q/(q-1)},
\end{align}
where $a_0$ is a new dimensionless parameter used to describe
mainly the shape of the spectrum in a low-$p_T$ region. Comparing
with $a_0=1$, $a_0>1$ means a lower spectrum and $a_0<1$ means a
higher spectrum. After introducing the index $a_0$, the tendency
of the spectrum in intermediate- and high-$p_T$ regions is also
changed due to the constraint of the normalization. Here, we would
like to point out that we made many attempts to find a suitable
function before this work. Some inadequacies always appeared with
the Tsallis distribution in different forms. Our various attempts
showed that the introduction of $a_0$ is necessary and useful. We
call Equation (2) the modified Tsallis-like distribution.

In the multi-source thermal model~\cite{5}, in high energy
collisions, we assume that two participant partons take part in
the formation of a given particle. The transverse momentum
$p_{T_1}$ ($p_{T_2}$) of the first (second) participant parton is
assumed to obey Equation~(2). That is, we have the probability density
function
\begin{align}
	f_i(p_{T_i},T)=\frac{1}{N_i}\frac{dN_i}{dp_{T_i}} = C
	m_{T_i}^{a_0}
	\bigg[1+\frac{(q-1)(m_{T_i}-\mu_i-m_{0_i})}{T}\bigg]^{-q/(q-1)},
\end{align}
where the subscript $i$ refers to 1 or 2, $m_{0_i}=0.31$ GeV/$c$
due to $u$ and $d$ quarks being involved mainly in AA collisions,
$N_i$ denotes the number of parton $i$, and $N_1=N_2=N$.

It should be noted that we have regarded $m_{0_i}$ as the
constituent masses which are the same for $u$ and $d$
quarks~\cite{5a}, but not the current masses, due to our
experiential choice. We do not need to consider an $s$ quark even for
a $\phi$ meson due to the fact that its formation are also from two
participant partons, $u$ and/or $d$ quarks, which are from the
projectile and target nuclei in AA collisions. Meanwhile, we do
not need to consider three constituent quarks for $p$; instead,
two participant partons from the projectile/target nuclei are
needed. That is, we consider only the projectile/target
participant quarks which can be regarded as two energy sources,
but not the constituent quarks of a given particle. Even for the
productions of leptons and jets~\cite{5b,5c}, the picture of two
participant quarks is applicable, nothing but two light (heavy)
quarks for the production of leptons (jets). Of course,
considering three constituent quarks for $p$ is another workable
picture~\cite{5b,5d} if the selected function is~appropriate.

The chemical potential of particles refers to the excess degree of
baryon number of positive matter relative to antimatter, so it
generally reflects the generation of particles at low
energy~\cite{6,7,8,9,10,11,12}. For baryons (mostly protons and
neutrons), the relationship between the collision energy
$\sqrt{s_{NN}}$ and chemical potential $\mu_B$ can be given by an
empirical formula:
\begin{align}
	\mu_B = \frac{1.3075}{1+0.288\sqrt{s_{NN}}}.
\end{align}

Among them, the units of $\mu_B$ and $\sqrt{s_{NN}}$ are
GeV~\cite{13,14,15,15a}. Since a proton or neutron is composed of
three $u/d$ quarks, we have $\mu_u = \mu_d = \mu_B/3$. Based on
different sets of data, the coefficients in Equation (4) may be
slightly invariant, though they are updated due to
Ref.~\cite{15a}.

In the Monte Carlo calculations, we need the discrete values of
$p_{T_1}$ and $p_{T_2}$. Let $R_1$ and $R_2$ are random numbers
distributed evenly in $[0,1]$. We have the expressions satisfied
by $p_{T_1}$ and $p_{T_2}$ to be

\begin{align}
	\int_0^{p_{T_1}} f_{p_{T_1}}(p'_{T_1},T)dp'_{T_1} <R_1<
	\int_0^{p_{T_1}+\delta p_{T_1}}
	f_{p_{T_1}}(p'_{T_1},T)dp'_{T_1},\\
	\int_0^{p_{T_2}} f_{p_{T_2}}(p'_{T_2},T)dp'_{T_2} <R_2<
	\int_0^{p_{T_2}+\delta p_{T_2}} f_{p_{T_2}}(p'_{T_2},T)dp'_{T_2},
\end{align}
where $\delta p_{T_1}$ and $\delta p_{T_2}$ are small amounts
added in $p_{T_1}$ and $p_{T_2}$, respectively.

Let $p_x$ ($p_y$) denote the $x$-component ($y$-component) of
particle's $p_T$, and $\phi_1$ ($\phi_2$) denote the isotropic
azimuth of the first (second) parton. We have the expressions for
$p_x$ and $p_y$ to~be
\begin{align}
	p_x = p_{T_1}\cos\phi_1+p_{T_2}\cos\phi_2=p_{T_1}\cos(2\pi
	R_3)+p_{T_2}\cos(2\pi R_4),\\
	p_y = p_{T_1}\sin\phi_1+p_{T_2}\sin\phi_2=p_{T_1}\sin(2\pi
	R_5)+p_{T_2}\sin(2\pi R_6),
\end{align}
where $R_{3,4,5,6}$ are random numbers distributed evenly in
$[0,1]$. After repeated calculations, we can obtain the
distribution of $p_T$ by the statistical method due to the fact
that \linebreak $p_T=\sqrt{p^2_x+p^2_y}$.

To perform a calculation based on Equations (5)--(8), we need a set of
concrete values of $R_1$ and $R_2$, respectively, from a
sub-program or special command on a random number in terms of a
given software such as the Matlab or Python. Then, we may search
suitable $p_{t_1}$ and $p_{t_2}$ that obey Equations (5) and (6) by a
code. In the statistics for repeated calculations, if the
distribution of $p_T$ is given by $dN/dp_T$, the joint
distribution of $y$ and $dp_T$ is simply given by $d^2N/dydp_T$,
which is obtained by $dN/dp_T$ being divided by $dy$, where $dy$
is a defined and small value at mid-$y$. In fact, in this work,
the minimum $dy=0.1$ and the maximum $dy=1$ which correspond to
$|y|<0.05$ and $|y|<0.5$ at mid-$y$, respectively.
\\

{\section{Results and Discussion}}

{\subsection{Comparison with the Data}}

Figure 1 shows the $p_T$ spectra, invariant yield $(1/2\pi
p_T)d^2N/dydp_T$, of $\pi^+$ (left panel) and $\pi^-$ (right
panel) produced at mid-$y$ or mid-$\eta$ in central AA collisions.
The experimental data (symbols) are from the HADES~\cite{22b},
E866~\cite{23}, E895~\cite{25}, STAR~\cite{28,29,30},
PHENIX~\cite{31,32}, and ALICE Collaborations~\cite{36,21}.
Different symbols represent the data at different energies (2.4,
2.7, 3.2, 3.84, 4.3, 4.85, 5.03, 7.7, 11.5, 14.5, 19.6, 27, 39,
62.4, 130, 200, and 2760 GeV), where the centralities for 2.4 GeV
and other energies are 0--10\% and 0--5\%, respectively. The solid
curves represent the result of our fit by using the Monte Carlo
method based on the modified Tsallis distribution. The dotted
curves represent a few examples from the Tsallis distribution for
comparisons. The energy 2760 GeV is for Pb-Pb collisions, while
the others are for Au-Au collisions. What we need to emphasize
here is that some data in the literature are given in the $m_T$
spectra which are converted by us to the $p_T$ spectra for the
unification. To see the data clearly and keep them from the
overlap, we multiply the data by the corresponding factors which
are listed in Table 1.

In the process of fitting the data, we used the least square
method to obtain the best parameters. The errors used to calculate
$\chi^2$ are obtained by the root-mean-square of statistical and
systematic errors. The parameters that minimize $\chi^2$ are the
best parameters. The parameter errors are obtained by the
statistical simulation method~\cite{18,19}. The collaborations,
free parameters ($T$, $q$, and $a_0$), normalization factor
($N_0$), and number of degree-of-freedom (ndof) are listed in
Table 1. From the comparisons between the solid and dotted curves
in Figure 1 and between the two sets of $\chi^2$ in Table 1, one
can see that the modified Tsallis distribution is better than the
Tsallis distribution in the fit. In view of these comparisons, we
give up using the Tsallis distribution in the fit for other data.

\renewcommand{\baselinestretch}{1.}

\begin{table*}[!htb]
{\small  Table 1. Values of $T$, $q$, $a_0$, $N_0$, $\chi^2$, and
ndof corresponding to the solid curves in Figure \ref{f1} in which $\pi^+$ (up
panel) and $\pi^-$ (down panel) data are measured by different
collaborations at different energies. Following the sets of parameters
for the three top energies, the sets of parameters corresponding to the
dotted curves are given. \label{t1}
\vspace{-.5cm}
\begin{center}
\begin{tabular} {ccccccccc}\\ \hline\hline Collab. &  $\sqrt{s_{NN}}$ (GeV) & Rapidity &Factor& $T$ (GeV) & $q$ & $a_0$ & $N_0$ &$\chi^2$/ndof\\
\hline
  HADES&        2.4 & $\lvert y \rvert <0.05$    &    5000 & $0.062\pm0.002$ & $1.080\pm0.002$ & $-0.55\pm0.02$ & $(2.1\pm0.1)\times 10^{-6}$ & $146/15$ \\
  E866 &        2.7 & $\lvert y \rvert <0.05$    &    0.01 & $0.130\pm0.003$ & $1.080\pm0.004$ & $-0.31\pm0.06$ & $12\pm1$                    & $150/19$ \\
  E866 &       3.32 & $\lvert y \rvert <0.05$    &    0.02 & $0.153\pm0.003$ & $1.110\pm0.004$ & $-0.18\pm0.06$ & $29\pm1$                    & $56/24$ \\
  E866 &       3.84 & $\lvert y \rvert <0.05$    &    0.05 & $0.162\pm0.002$ & $1.120\pm0.003$ & $-0.01\pm0.02$ & $39\pm2$                    & $48/19$ \\
  E866 &        4.3 & $\lvert y \rvert <0.05$    &     0.1 & $0.164\pm0.002$ & $1.122\pm0.005$ & $-0.05\pm0.02$ & $48\pm2$                    & $22/16$ \\
  E866 &       4.85 & $\lvert y \rvert <0.05$    &     0.2 & $0.170\pm0.003$ & $1.126\pm0.006$ & $-0.05\pm0.04$ & $52\pm1$                    & $26/16$ \\
  E802 &       5.03 & $\lvert y \rvert <0.2$     &     0.5 & $0.176\pm0.002$ & $1.130\pm0.006$ & $-0.07\pm0.03$ & $55\pm1$                    & $104/30$ \\
  STAR &        7.7 & $\lvert y \rvert <0.1$     &     0.8 & $0.180\pm0.001$ & $1.130\pm0.001$ & $-0.07\pm0.01$ & $98\pm2$                    & $41/22$ \\
  STAR &       11.5 & $\lvert y \rvert <0.1$     &       1 & $0.180\pm0.002$ & $1.140\pm0.002$ & $-0.10\pm0.02$ & $131\pm4$                   & $15/22$ \\
  STAR &       14.5 & $\lvert y \rvert <0.1$     &       2 & $0.180\pm0.002$ & $1.142\pm0.002$ & $-0.10\pm0.03$ & $161\pm2$                   & $3/24$ \\
  STAR &       19.6 & $\lvert y \rvert <0.1$     &       5 & $0.184\pm0.002$ & $1.146\pm0.004$ & $-0.10\pm0.04$ & $169\pm2$                   & $17/22$ \\
  STAR &         27 & $\lvert y \rvert <0.1$     &      10 & $0.186\pm0.001$ & $1.146\pm0.002$ & $-0.10\pm0.01$ & $179\pm3$                   & $9/21$ \\
  STAR &         39 & $\lvert y \rvert <0.1$     &      20 & $0.189\pm0.002$ & $1.146\pm0.003$ & $-0.10\pm0.03$ & $189\pm5$                   & $4/22$ \\
  STAR &       62.4 & $\lvert y \rvert <0.1$     &      50 & $0.188\pm0.003$ & $1.144\pm0.004$ & $-0.11\pm0.02$ & $239\pm2$                   & $1/6$ \\
PHENIX &        130 & $\lvert \eta \rvert <0.35$ &     100 & $0.183\pm0.005$ & $1.140\pm0.005$ & $-0.15\pm0.03$ & $253\pm37$                  & $43/10$ \\
       &            &                            &         & $0.153\pm0.004$ & $1.080\pm0.003$ &                & $253\pm37$                  & $192/11$ \\
PHENIX &        200 & $\lvert \eta \rvert <0.35$ &     400 & $0.188\pm0.003$ & $1.143\pm0.001$ & $-0.05\pm0.03$ & $306\pm12$                  & $108/24$ \\
       &            &                            &         & $0.178\pm0.005$ & $1.043\pm0.002$ &                & $304\pm12$                  & $561/25$ \\
 ALICE &       2760 & $\lvert y \rvert <0.5$     &     500 & $0.227\pm0.002$ & $1.178\pm0.002$ & $-0.01\pm0.01$ & $750\pm27$                  & $44/37$ \\
       &            &                            &         & $0.187\pm0.003$ & $1.087\pm0.001$ &                & $750\pm26$                  & $234/38$ \\
\hline
 HADES &        2.4 & $\lvert y \rvert <0.05$     &   3000  & $0.078\pm0.002$ & $1.049\pm0.002$ & $-0.56\pm0.02$ & $(9.1\pm0.1)\times 10^{-5}$ & $184/29$ \\
  E895 &        2.7 & $\lvert y \rvert <0.05$     &    0.01 & $0.110\pm0.003$ & $1.060\pm0.004$ & $-0.41\pm0.06$ & $19\pm2$                    & $395/26$ \\
  E895 &       3.32 & $\lvert y \rvert <0.05$     &    0.02 & $0.143\pm0.003$ & $1.104\pm0.004$ & $-0.23\pm0.06$ & $38\pm1$                    & $389/36$ \\
  E895 &       3.84 & $\lvert y \rvert <0.05$     &    0.05 & $0.150\pm0.002$ & $1.120\pm0.003$ & $-0.04\pm0.02$ & $48\pm2$                    & $189/36$ \\
  E895 &        4.3 & $\lvert y \rvert <0.05$     &     0.1 & $0.155\pm0.002$ & $1.102\pm0.005$ & $-0.10\pm0.02$ & $62\pm2$                    & $242/36$ \\
  E802 &       5.03 & $0<y <0.4$                  &     0.5 & $0.170\pm0.002$ & $1.130\pm0.006$ & $-0.07\pm0.03$ & $64\pm1$                    & $137/29$ \\
  STAR &        7.7 & $\lvert y \rvert <0.1$      &     0.8 & $0.180\pm0.001$ & $1.128\pm0.001$ & $-0.07\pm0.01$ & $105\pm3$                   & $75/22$ \\
  STAR &       11.5 & $\lvert y \rvert <0.1$      &       1 & $0.177\pm0.002$ & $1.140\pm0.002$ & $-0.10\pm0.02$ & $137\pm2$                   & $14/22$ \\
  STAR &       14.5 & $\lvert y \rvert <0.1$      &       2 & $0.180\pm0.002$ & $1.142\pm0.002$ & $-0.10\pm0.03$ & $158\pm3$                   & $5/24$ \\
  STAR &       19.6 & $\lvert y \rvert <0.1$      &       5 & $0.184\pm0.002$ & $1.146\pm0.004$ & $-0.10\pm0.04$ & $170\pm3$                   & $17/21$ \\
  STAR &         27 & $\lvert y \rvert <0.1$      &      10 & $0.186\pm0.001$ & $1.146\pm0.002$ & $-0.10\pm0.01$ & $179\pm3$                   & $12/22$ \\
  STAR &         39 & $\lvert y \rvert <0.1$      &      20 & $0.189\pm0.004$ & $1.146\pm0.003$ & $-0.10\pm0.02$ & $189\pm3$                   & $6/22$ \\
  STAR &       62.4 & $\lvert y \rvert <0.1$      &      50 & $0.189\pm0.003$ & $1.144\pm0.004$ & $-0.11\pm0.02$ & $241\pm10$                  & $1/6$ \\
PHENIX &        130 & $\lvert \eta \rvert <0.35$  &     100 & $0.186\pm0.005$ & $1.149\pm0.005$ & $-0.14\pm0.03$ & $231\pm27$                  & $69/10$ \\
       &            &                             &         & $0.153\pm0.004$ & $1.080\pm0.002$ &                & $231\pm27$                  & $192/11$ \\
PHENIX &        200 & $\lvert \eta \rvert <0.35$  &     400 & $0.192\pm0.003$ & $1.143\pm0.001$ & $-0.05\pm0.03$ & $297\pm18$                  & $114/24$ \\
       &            &                             &         & $0.178\pm0.004$ & $1.043\pm0.001$ &                & $297\pm17$                  & $478/25$ \\
 ALICE &       2760 & $\lvert y \rvert <0.5$      &     500 & $0.227\pm0.002$ & $1.178\pm0.002$ & $-0.01\pm0.01$ & $738\pm27$                  & $54/37$ \\
       &            &                             &         & $0.187\pm0.003$ & $1.043\pm0.002$ &                & $738\pm26$                  & $223/38$ \\
\hline
\end{tabular}%
\end{center}}
\end{table*}

Figure 2 is similar to Figure 1, but it shows the invariant yield
of $K^+$ (left panel) and $K^-$ (right panel) produced at mid-$y$
or mid-$\eta$ in central Au-Au and Pb-Pb collisions. The data are
from the KaoS~\cite{22a}, HADES~\cite{22b}, E866~\cite{23},
E802~\cite{26}, STAR~\cite{28,29,30}, PHENIX~\cite{31,32}, and
ALICE Collaborations~\cite{36,21} over an energy range from 2.16
to 2760 GeV, where the centralities for 2.4 GeV and other energies
are 0--40\% and 0--5\%, respectively (0--5.4\% for 2.16, 2.24,
2.32, and 2.52 GeV, which is not marked in the panels). To see the
data clearly and keep them from the overlap, we multiply the data
by the corresponding factors. Similarly, the collaborations, $T$,
$q$, $a_0$, $N_0$, and ndof are listed in Table 2 with the
factors.

\begin{figure*}[!htb]
\begin{center}
\includegraphics[width=16.7cm]{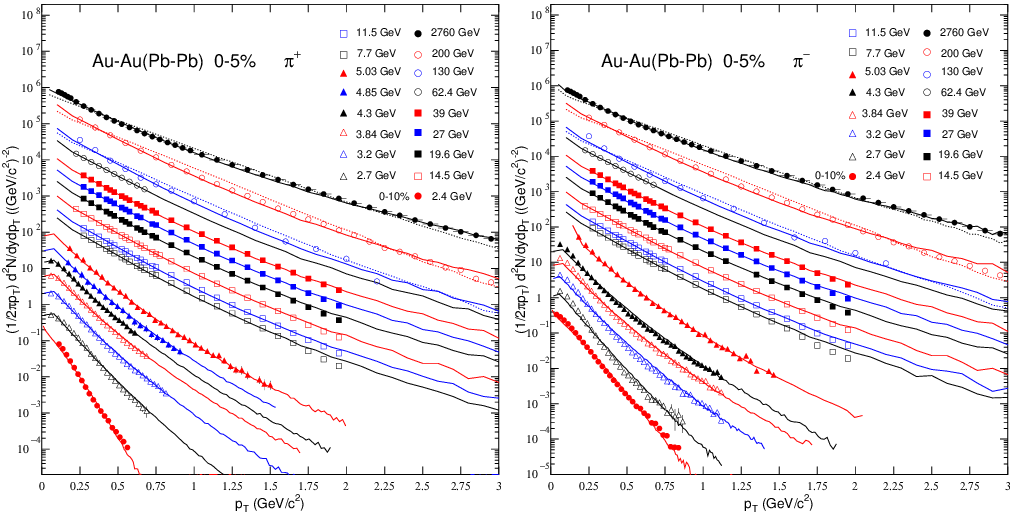}
\end{center}
{\small Fig. 1. Invariant yield of $\pi^+$ (left panel) and $\pi^-$
(right panel) produced at mid-$y$ or mid-$\eta$ in central Au-Au
and Pb-Pb collisions. The experimental data (symbols) are from
the HADES~\cite{22b}, E866~\cite{23}, E895~\cite{25}, STAR~\cite{28,29,30},
PHENIX~\cite{31,32}, and ALICE Collaborations~\cite{36,21} in the energy
range of 2.4--2760~GeV. Different symbols represent the data at
different energies. The energy 2760~GeV is for Pb-Pb collisions,
while the others are for Au-Au collisions. The solid curves represent
the result of our fit by using the Monte Carlo method based on the
modified Tsallis-like distribution. The dotted curves represent
a few examples from the Tsallis distribution for comparisons. The factors
multiplied to distinguish the data are listed in Table 1.\label{f1}}
\end{figure*}

\begin{figure*}[!htb]
\begin{center}
\includegraphics[width=16.7cm]{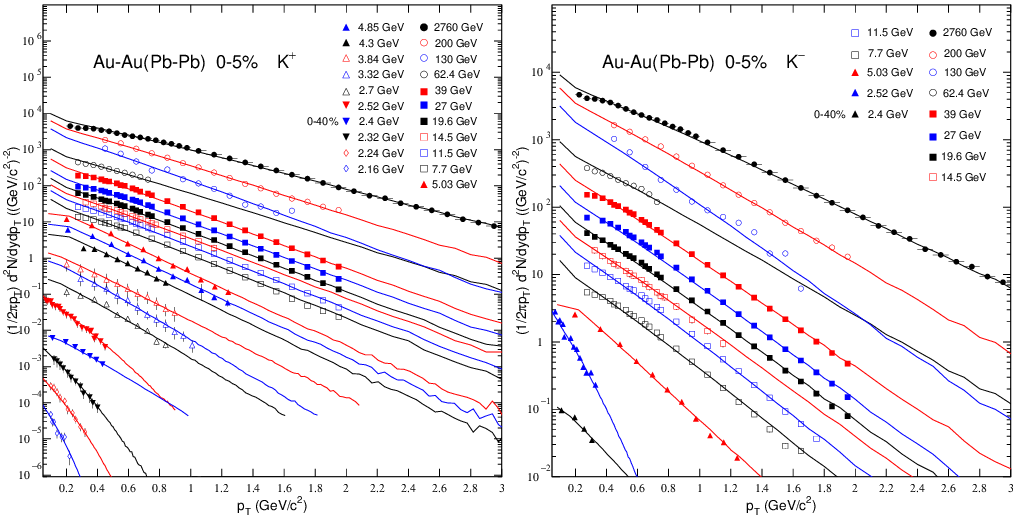}
\end{center}
{\small Fig. 2. Same as Figure 1, but showing the invariant yield
of $K^+$ (left panel) and $K^-$ (right panel). The data are from
the KaoS~\cite{22a}, HADES~\cite{22b}, E866~\cite{23}, E802~\cite{26},
STAR~\cite{28,29,30}, PHENIX~\cite{31,32}, and ALICE Collaborations~\cite{36,21}
in the energy range of 2.16--2760 GeV. Only the solid curves are available.
The factors multiplied to distinguish the data are listed in Table 2.\label{f2}}
\end{figure*}

\renewcommand{\baselinestretch}{1.}
\begin{table*}[!htb]
{\small Table 2. Values of $T$, $q$, $a_0$, $N_0$,
$\chi^2$, and ndof corresponding to the curves in Figure 2 in
which $K^+$ (up panel) and $K^-$ (down panel) data are measured by
different collaborations at different energies. In one case, ndof
is less than 1, which is denoted by $-$ in the table, and the
corresponding curve is obtained by an extrapolation. \label{t2}
\vspace{-.5cm}
\begin{center}
\begin{tabular}{ccccccccc}\\ \hline\hline Collab. &$\sqrt{s_{NN}}$ (GeV) & Rapidity &Factor& $T$ (GeV) & $q$ & $a_0$ &$N_0$ & $\chi^2$/ndof \\
\hline
  KaoS &       2.16 & $\lvert y \rvert <0.5$     &        0.1        & $0.020\pm0.002$ & $1.002\pm0.003$ & $-0.56\pm0.05$ & $(4.3\pm0.1)\times 10^{-5}$ & $4/3$ \\
  KaoS &       2.24 & $\lvert y \rvert <0.5$     &        0.05       & $0.037\pm0.002$ & $1.004\pm0.003$ & $-0.46\pm0.05$ & $(7.3\pm0.1)\times 10^{-4}$ & $2/6$ \\
  KaoS &       2.32 & $\lvert y \rvert <0.5$     &        0.1        & $0.052\pm0.002$ & $1.004\pm0.003$ & $-0.46\pm0.05$ & $(3.3\pm0.1)\times 10^{-3}$ & $1/8$ \\
  HADES&        2.4 & $\lvert y \rvert <0.1$     & $5\times 10^{10}$ & $0.154\pm0.004$ & $1.010\pm0.005$ & $0.56\pm0.05$  & $(2.1\pm0.1)\times 10^{-13}$& $7/3$ \\
  KaoS &       2.52 & $\lvert y \rvert <0.5$     &        0.3        & $0.089\pm0.002$ & $1.014\pm0.003$ & $-0.36\pm0.05$ & $0.5\pm0.3$                 & $4/16$ \\
  E866 &        2.7 & $\lvert y \rvert<0.23$     &          1        & $0.159\pm0.004$ & $1.010\pm0.005$ & $0.56\pm0.05$  & $0.40\pm0.03$               & $67/6$ \\
  E866 &       3.32 & $\lvert y \rvert<0.29$     &        0.5        & $0.160\pm0.002$ & $1.010\pm0.005$ & $0.53\pm0.02$  & $2.60\pm0.09$               & $45/8$ \\
  E866 &       3.84 & $\lvert y \rvert<0.05$     &        0.5        & $0.189\pm0.002$ & $1.029\pm0.002$ & $0.69\pm0.01$  & $5.40\pm0.09$               & $34/7$ \\
  E866 &        4.3 & $\lvert y \rvert<0.05$     &        1.5        & $0.195\pm0.001$ & $1.031\pm0.003$ & $0.72\pm0.02$  & $8.8\pm0.2$                 & $25/5$ \\
  E866 &       4.85 & $\lvert y \rvert <0.2$     &          2        & $0.198\pm0.005$ & $1.033\pm0.003$ & $0.72\pm0.02$  & $13\pm1$                    & $31/7$ \\
  E802 &       5.03 & $0<y <0.4$                 &          4        & $0.199\pm0.002$ & $1.034\pm0.001$ & $0.73\pm0.03$  & $13\pm1$                    & $32/7$ \\
  STAR &        7.7 & $\lvert y \rvert <0.1$     &          1        & $0.197\pm0.002$ & $1.033\pm0.003$ & $0.72\pm0.05$  & $23\pm1$                    & $40/19$ \\
  STAR &       11.5 & $\lvert y \rvert <0.1$     &        1.5        & $0.197\pm0.001$ & $1.033\pm0.002$ & $0.72\pm0.01$  & $27\pm1$                    & $32/21$ \\
  STAR &       14.5 & $\lvert y \rvert <0.1$     &          2        & $0.203\pm0.002$ & $1.036\pm0.004$ & $0.74\pm0.03$  & $31\pm1$                    & $1/14$ \\
  STAR &       19.6 & $\lvert y \rvert <0.1$     &          3        & $0.203\pm0.001$ & $1.036\pm0.002$ & $0.74\pm0.03$  & $31\pm1$                    & $16/22$ \\
  STAR &         27 & $\lvert y \rvert <0.1$     &          5        & $0.207\pm0.002$ & $1.037\pm0.003$ & $0.79\pm0.03$  & $32\pm1$                    & $39/22$ \\
  STAR &         39 & $\lvert y \rvert <0.1$     &         10        & $0.207\pm0.002$ & $1.037\pm0.003$ & $0.79\pm0.04$  & $34\pm1$                    & $29/22$ \\
  STAR &       62.4 & $\lvert y \rvert <0.1$     &         20        & $0.229\pm0.002$ & $1.042\pm0.001$ & $0.99\pm0.01$  & $42\pm1$                    & $8/6$ \\
PHENIX &        130 & $\lvert \eta \rvert <0.35$ &         50        & $0.207\pm0.005$ & $1.037\pm0.006$ & $0.79\pm0.05$  & $46\pm6$                    & $11/9$ \\
PHENIX &        200 & $\lvert \eta \rvert <0.35$ &        100        & $0.227\pm0.003$ & $1.042\pm0.002$ & $0.99\pm0.05$  & $49\pm3$                    & $15/12$ \\
 ALICE &       2760 & $\lvert y \rvert <0.5$     &        100        & $0.242\pm0.002$ & $1.062\pm0.002$ & $1.39\pm0.02$  & $114\pm4$                   & $25/32$ \\
\hline
 HADES &        2.4 & $\lvert y \rvert <0.05$    &$5\times 10^{13}$ & $0.154\pm0.004$ & $1.010\pm0.005$ & $0.56\pm0.05$ & $(3.2\pm0.1)\times 10^{-15}$ & $4/-$ \\
 KaoS  &       2.52 & $\lvert y \rvert <0.5$     &              500 & $0.063\pm0.002$ & $1.010\pm0.003$ & $-0.46\pm0.05$& $(6.6\pm0.1)\times 10^{-3}$  & $11/79$ \\
  E802 &       5.03 & $\lvert y \rvert <0.1$     &                4 & $0.193\pm0.002$ & $1.031\pm0.001$ & $0.66\pm0.03$ & $2.5\pm0.0$                  & $34/37$ \\
  STAR &        7.7 & $\lvert y \rvert <0.1$     &                1 & $0.190\pm0.002$ & $1.030\pm0.003$ & $0.68\pm0.05$ & $8.3\pm0.2$                  & $43/19$ \\
  STAR &       11.5 & $\lvert y \rvert <0.1$     &              1.5 & $0.190\pm0.001$ & $1.030\pm0.002$ & $0.70\pm0.01$ & $13\pm1$                     & $27/19$ \\
  STAR &       14.5 & $\lvert y \rvert <0.1$     &                2 & $0.200\pm0.002$ & $1.035\pm0.004$ & $0.72\pm0.03$ & $18\pm1$                     & $1/14$ \\
  STAR &       19.6 & $\lvert y \rvert <0.1$     &                3 & $0.201\pm0.001$ & $1.034\pm0.002$ & $0.73\pm0.03$ & $20\pm2$                     & $15/22$ \\
  STAR &         27 & $\lvert y \rvert <0.1$     &                5 & $0.200\pm0.002$ & $1.032\pm0.003$ & $0.75\pm0.03$ & $24\pm1$                     & $30/20$ \\
  STAR &         39 & $\lvert y \rvert <0.1$     &               10 & $0.208\pm0.002$ & $1.037\pm0.003$ & $0.79\pm0.04$ & $27\pm1$                     & $20/22$ \\
  STAR &       62.4 & $\lvert y \rvert <0.1$     &               20 & $0.230\pm0.002$ & $1.042\pm0.001$ & $0.99\pm0.01$ & $38\pm1$                     & $22/6$ \\
PHENIX &        130 & $\lvert \eta \rvert <0.35$ &               50 & $0.204\pm0.005$ & $1.033\pm0.006$ & $0.76\pm0.05$ & $38\pm6$                     & $9/9$ \\
PHENIX &        200 & $\lvert \eta \rvert <0.35$ &              100 & $0.227\pm0.003$ & $1.042\pm0.002$ & $0.99\pm0.05$ & $46\pm5$                     & $11/12$ \\
 ALICE &       2760 & $\lvert y \rvert <0.5$     &              100 & $0.247\pm0.002$ & $1.052\pm0.002$ & $1.39\pm0.02$ & $110\pm4$                    & $15/32$ \\
\hline
\end{tabular}%
\end{center}}
\end{table*}

Figure 3 is similar to Figures 1 and 2, but it shows the invariant
yield of $p$ (left panel) and $\bar p$ (right panel) produced at
mid-$y$ or mid-$\eta$ in 0--5\% Au-Au and Pb-Pb collisions. The
data are from the E895~\cite{24}, E802~\cite{27},
STAR~\cite{28,29,30}, PHENIX~\cite{31,32}, and ALICE
Collaborations~\cite{36,21} in the energy range of 2.7--2760 GeV.
To see the data clearly and keep them from the overlap, we
multiply the data by the corresponding factors. Similarly, the
collaborations, $T$, $q$, $a_0$, $N_0$, and ndof are listed in
Table 3 with the factors.

Figure 4 is similar to Figures 3, but it shows the invariant
yield of $\phi$ produced at mid-$y$ in central Au-Au and Pb-Pb
collisions. The data are from the HADES~\cite{22c},
STAR~\cite{37,38}, and ALICE Collaborations~\cite{39}. The
energies are 2.4, 7.7, 11.5, 19.6, 27, 39, 62.4, 130, 200, and
2760 GeV, where the centralities for 2.4, 62.4, and 130 GeV are
0--40\%, 0--20\%, and 0--11\%, respectively, and for other
energies are 0--5\%. To see the data clearly and keep them from
the overlap, we multiply the data by the corresponding factors.
Similarly, the collaborations, $T$, $q$, $a_0$, $N_0$, and ndof
are listed in Table 4 with the factors.

\begin{figure*}[!htb]
\begin{center}
\includegraphics[width=16.7cm]{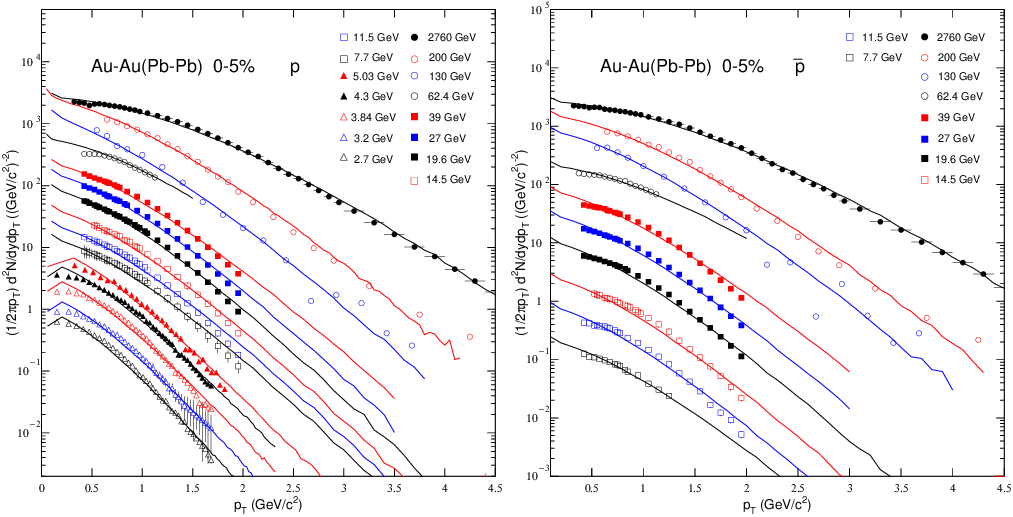}
\end{center}
{\small Fig. 3. Same as Figures 1 and 2, but showing the invariant
yield of $p$ (left panel) and $\bar p$ (right panel). The data are
from the E895~\cite{24}, E802~\cite{27}, STAR~\cite{28,29,30},
PHENIX~\cite{31,32}, and ALICE Collaborations~\cite{36,21} in the
energy range of 2.7--2760 GeV. The factors multiplied to distinguish
the data are listed in Table 3.\label{f3}}
\end{figure*}

\begin{figure*}[!htb]
\begin{center}
\includegraphics[width=10.8cm]{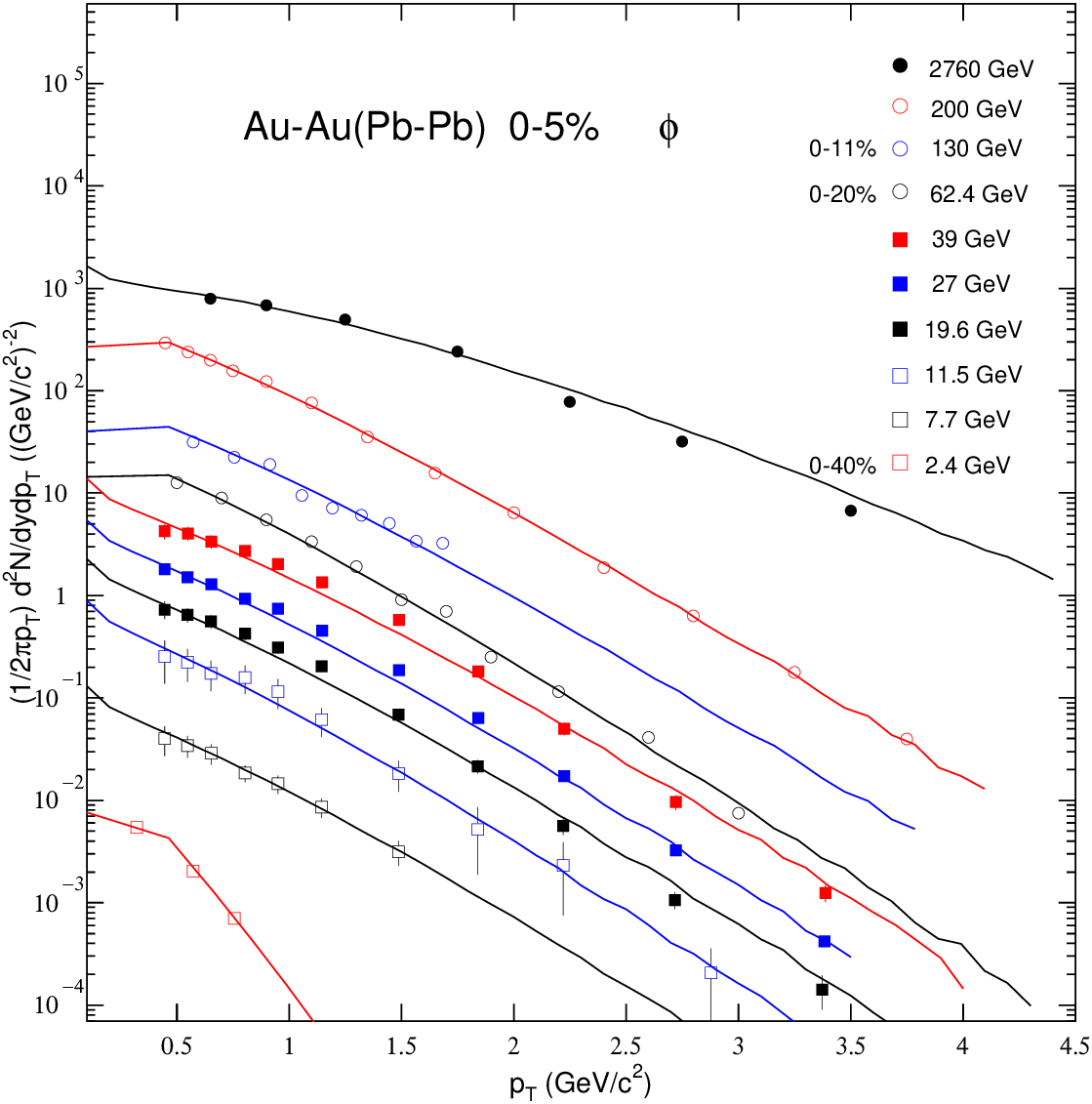}
\end{center}
{\small Fig. 4. Same as Figures 1--3, but showing the invariant
yield of $\phi$. The data are from the HADES~\cite{22c},
STAR~\cite{37,38}, and ALICE Collaborations~\cite{39} in the
energy range of 2.4--2760 GeV. The factors multiplied to distinguish
the data are listed in Table 4.\label{f4}}
\end{figure*}

\begin{table*}[!htb]
{\small Table 3. Values of $T$, $q$, $a_0$, $N_0$, $\chi^2$, and
ndof corresponding to the curves in Figure 3 in which $p$ (up
panel) and $\bar p$ (down panel) data are measured by different
collaborations at different energies.
\vspace{-.5cm}
\begin{center}
\begin{tabular} {ccccccccc}\\ \hline\hline Collab. & $\sqrt{s_{NN}}$ (GeV) & Rapidity &Factor& $T$ (GeV) & $q$ &$a_0$ & $N_0$ & $\chi^2$/ndof \\
\hline
				E895 &        2.7 & $\lvert y \rvert <0.05$    &   0.1 & $0.180\pm0.003$ & $1.005\pm0.004$ & $1.35\pm0.08$ & $75\pm6$      & $256/36$ \\
				E895 &       3.32 & $\lvert y \rvert <0.05$    &   0.2 & $0.187\pm0.003$ & $1.008\pm0.005$ & $1.38\pm0.08$ & $71\pm5$      & $217/36$ \\
				E895 &       3.84 & $\lvert y \rvert <0.34$    &   0.5 & $0.193\pm0.002$ & $1.008\pm0.002$ & $1.38\pm0.05$ & $64\pm6$      & $309/36$ \\
				E895 &        4.3 & $\lvert y \rvert <0.37$    &     1 & $0.199\pm0.004$ & $1.011\pm0.002$ & $1.42\pm0.06$ & $60\pm4$      & $236/36$ \\
				E802 &       5.03 & $\lvert y \rvert <0.1$     &   1.5 & $0.199\pm0.001$ & $1.013\pm0.002$ & $1.44\pm0.06$ & $69\pm2$      & $465/25$ \\
				STAR &        7.7 & $\lvert y \rvert <0.1$     &   0.5 & $0.204\pm0.001$ & $1.010\pm0.003$ & $1.44\pm0.02$ & $56\pm1$      & $48/25$ \\
				STAR &       11.5 & $\lvert y \rvert <0.1$     &     1 & $0.204\pm0.001$ & $1.010\pm0.002$ & $1.44\pm0.04$ & $46\pm2$      & $61/24$ \\
				STAR &       14.5 & $\lvert y \rvert <0.1$     &     2 & $0.204\pm0.001$ & $1.010\pm0.003$ & $1.44\pm0.04$ & $42\pm1$      & $6/21$ \\
				STAR &       19.6 & $\lvert y \rvert <0.1$     &     5 & $0.205\pm0.002$ & $1.010\pm0.002$ & $1.44\pm0.06$ & $36\pm1$      & $24/25$ \\
				STAR &         27 & $\lvert y \rvert <0.1$     &    10 & $0.209\pm0.002$ & $1.010\pm0.002$ & $1.44\pm0.07$ & $33\pm1$      & $15/19$ \\
				STAR &         39 & $\lvert y \rvert <0.1$     &    20 & $0.219\pm0.001$ & $1.010\pm0.001$ & $1.47\pm0.02$ & $27\pm1$      & $17/18$ \\
				STAR &       62.4 & $\lvert y \rvert <0.1$     &    50 & $0.239\pm0.002$ & $1.009\pm0.004$ & $1.87\pm0.04$ & $34\pm1$      & $46/11$ \\
		      PHENIX &        130 & $\lvert \eta \rvert <0.35$ &   100 & $0.223\pm0.002$ & $1.020\pm0.002$ & $1.51\pm0.05$ & $28\pm1$      & $629/13$ \\
			  PHENIX &        200 & $\lvert \eta \rvert <0.35$ &   400 & $0.233\pm0.001$ & $1.003\pm0.001$ & $1.71\pm0.02$ & $17\pm1$      & $83/18$ \\
			   ALICE &       2760 & $\lvert y \rvert <0.5$     &   500 & $0.263\pm0.001$ & $1.003\pm0.002$ & $2.21\pm0.05$ & $33\pm2$      & $102/38$ \\
				\hline
				STAR &        7.7 & $\lvert y \rvert <0.1$     &     1 & $0.208\pm0.001$ & $1.013\pm0.003$ & $1.47\pm0.02$ & $0.40\pm0.01$ & $5/11$ \\
				STAR &       11.5 & $\lvert y \rvert <0.1$     &     1 & $0.200\pm0.001$ & $1.010\pm0.002$ & $1.40\pm0.04$ & $1.50\pm0.03$ & $57/19$ \\
				STAR &       14.5 & $\lvert y \rvert <0.1$     &     2 & $0.201\pm0.001$ & $1.010\pm0.003$ & $1.44\pm0.04$ & $2.5\pm0.2$   & $8/21$ \\
				STAR &       19.6 & $\lvert y \rvert <0.1$     &     5 & $0.204\pm0.002$ & $1.010\pm0.002$ & $1.44\pm0.06$ & $4.2\pm0.1$   & $31/18$ \\
				STAR &         27 & $\lvert y \rvert <0.1$     &    10 & $0.209\pm0.002$ & $1.010\pm0.002$ & $1.44\pm0.07$ & $6.4\pm0.1$   & $22/18$ \\
				STAR &         39 & $\lvert y \rvert <0.1$     &    20 & $0.217\pm0.001$ & $1.010\pm0.001$ & $1.47\pm0.02$ & $9.0\pm0.1$   & $320/19$ \\
				STAR &       62.4 & $\lvert y \rvert <0.1$     &    50 & $0.239\pm0.002$ & $1.009\pm0.004$ & $1.87\pm0.04$ & $17\pm1$      & $18/12$ \\
			  PHENIX &        130 & $\lvert \eta \rvert <0.35$ &   100 & $0.233\pm0.002$ & $1.007\pm0.002$ & $1.51\pm0.05$ & $21\pm1$      & $23/13$ \\
			  PHENIX &        200 & $\lvert \eta \rvert <0.35$ &   400 & $0.233\pm0.001$ & $1.003\pm0.001$ & $1.71\pm0.02$ & $12\pm1$      & $60/18$ \\
			   ALICE &       2760 & $\lvert y \rvert <0.5$     &   500 & $0.263\pm0.001$ & $1.003\pm0.002$ & $2.21\pm0.05$ & $33\pm2$      & $111/38$ \\
				\hline
\end{tabular}%
\end{center}}
\end{table*}

\begin{table*}[!htb]
{\small Table 4. Values of $T$, $q$, $a_0$, $N_0$, $\chi^2$, and
ndof corresponding to the curves in Figure \ref{f4} in which $\phi$ data
are measured by different collaborations at different energies.
In one case, ndof is less than 1, which is denoted by $-$ in the table,
and the corresponding curve is obtained by an extrapolation.
\vspace{-.5cm}
\begin{center}
\begin{tabular} {ccccccccc}\\ \hline\hline	Collab. & $\sqrt{s_{NN}}$ (GeV) & Rapidity &Factor& $T$ (GeV) & $q$& $a_0$ & $N_0$ & $\chi^2$/ndof\\
	\hline
		   HADES &        2.4 & $\lvert y \rvert <0.1$ & $1\times 10^{14}$ & $0.124\pm0.001$ & $1.002\pm0.001$ & $0.60\pm0.06$ & $(6.8\pm0.1)\times 10^{-16}$ & $3$/- \\
			STAR &        7.7 & $\lvert y \rvert <0.5$ &               0.1 & $0.244\pm0.002$ & $1.009\pm0.001$ & $1.40\pm0.03$ & $1.4\pm0.1$                  & $10/3$ \\
			STAR &       11.5 & $\lvert y \rvert <0.5$ &               0.5 & $0.242\pm0.001$ & $1.006\pm0.001$ & $1.30\pm0.01$ & $1.8\pm0.1$                  & $7/6$ \\
			STAR &       19.6 & $\lvert y \rvert <0.5$ &                 1 & $0.247\pm0.001$ & $1.006\pm0.001$ & $1.40\pm0.02$ & $2.5\pm0.2$                  & $15/7$ \\
			STAR &         27 & $\lvert y \rvert <0.5$ &                 2 & $0.247\pm0.001$ & $1.006\pm0.001$ & $1.40\pm0.06$ & $3.0\pm0.4$                  & $10/8$ \\
			STAR &         39 & $\lvert y \rvert <0.5$ &                 5 & $0.251\pm0.002$ & $1.006\pm0.001$ & $1.50\pm0.07$ & $3.3\pm0.3$                  & $1/8$ \\
			STAR &       62.4 & $\lvert y \rvert <0.5$ &               100 & $0.251\pm0.003$ & $1.006\pm0.001$ & $1.80\pm0.06$ & $3.8\pm0.4$                  & $10/7$ \\
			STAR &        130 & $\lvert y \rvert <0.5$ &               200 & $0.261\pm0.003$ & $1.006\pm0.001$ & $2.00\pm0.07$ & $6.3\pm0.2$                  & $5/5$ \\
			STAR &        200 & $\lvert y \rvert <0.5$ &              1000 & $0.261\pm0.001$ & $1.006\pm0.001$ & $2.00\pm0.06$ & $8.4\pm0.5$                  & $8/10$ \\
			ALICE &      2760 & $\lvert y \rvert <0.5$ &               500 & $0.281\pm0.001$ & $1.012\pm0.001$ & $2.90\pm0.06$ & $14\pm1$                     & $1/4$ \\
			\hline
\end{tabular}%
\end{center}}
\end{table*}
\renewcommand{\baselinestretch}{1.15}

One can see from Figures 1--4 and Tables 1--4 that our results by
the Monte Carlo method describe approximately the tendency of the
considered experimental data. In our work, due to the narrow range
of $p_T$ spectra being used, we have considered only two
participant partons and one component (temperature). It is natural
and easy that we can extend this work to three or more participant
partons and two or more components (temperatures) if needed. In
the Monte Carlo calculations, adding the contributions of more
participant partons means increasing the number of items in Equations 
(7) and (8), while adding more components (temperatures) means
increasing new Equations (7) and (8) with different temperatures in the
calculations and different proportions in the statistics.
\\

{\subsection{Tendency of Parameters and Discussion}}

In order to study the change trend of parameters, Figure 5 shows
the dependences of (a) $T$, (b) $q$, (c) $a_0$, and (d) $N_0$ on
collision energy $\sqrt{s_{NN}}$. In the figure, the squares,
circles, triangles, and crosses represent the results from the
$\pi^{\pm}$, $K^{\pm}$, $p(\bar p)$, and $\phi$ spectra,
respectively. The closed symbols indicate the positive particles,
and the open ones indicate the negative particles. From Figure \ref{f5},
one can see that, as $\sqrt{s_{NN}}$ increases, the parameter $T$
increases quickly from 2.16 to 5 GeV and then slightly from 5 to
2760 GeV in the results from $\pi^{\pm}$ and $K^{\pm}$ spectra.
The parameter $q$ for $\pi^{\pm}$ increases quickly and then
slowly around 5 GeV, and for $K^{\pm}$ and $p(\bar p)$ shows a
slight increase. The parameter $a_0$ shows a slight increase or
remains almost unchanged in most cases. The parameter $N_0$
decreases for $p$ and increases quickly and then slowly around 5
GeV for $\pi^{\pm}$ and $K^{\pm}$. These parameters also show the
particle mass dependent. That is, with the increase of particle
mass, $T$ and $a_0$ increase and $q$ and $N_0$ decrease in most
cases.

\begin{figure*}[!htb]
\begin{center}
\includegraphics[width=15.1cm]{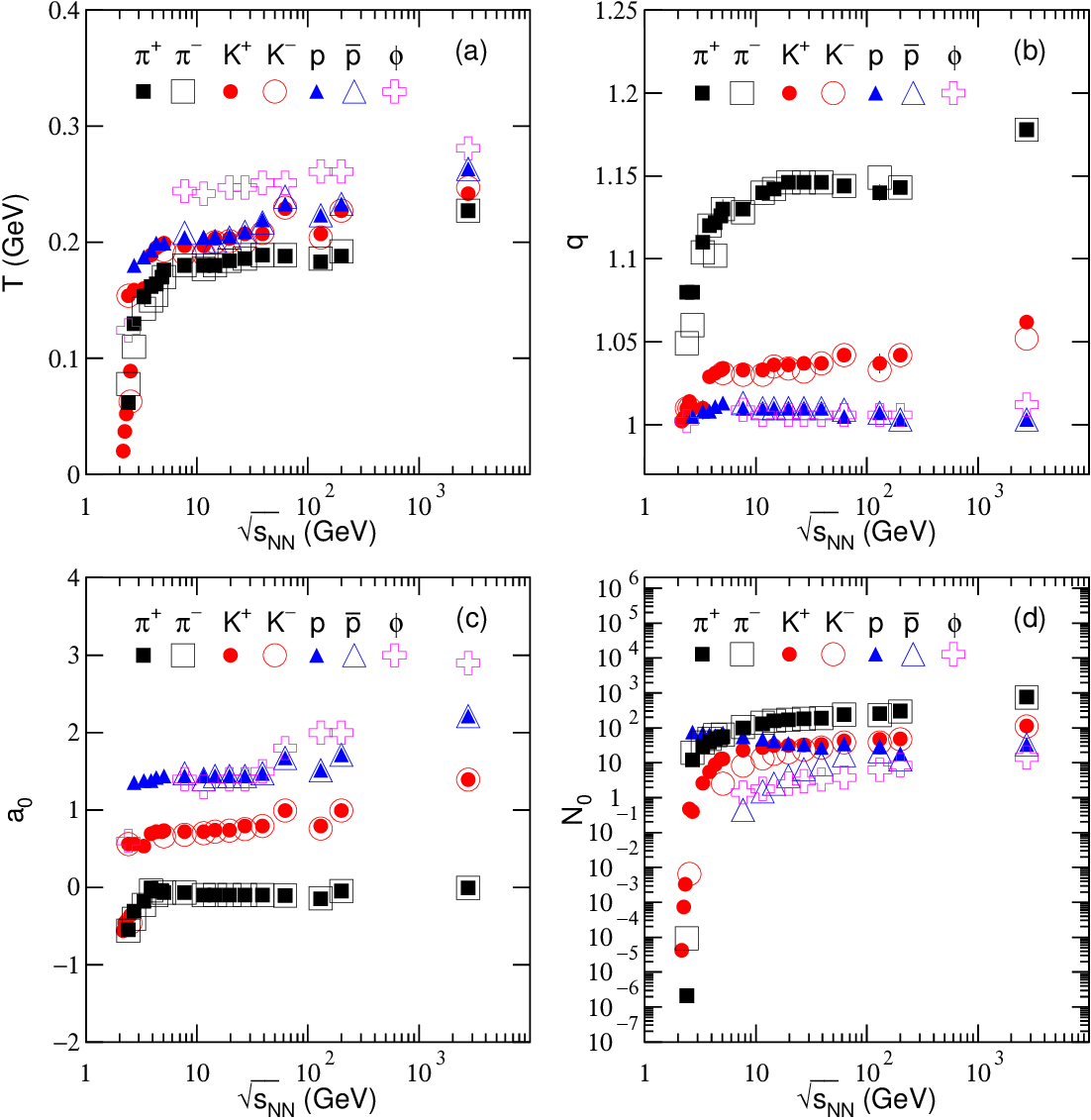}
\end{center}
{\small Fig. 5. Dependences of (\textbf{a}) $T$, (\textbf{b}) $q$, (\textbf{c}) $a_0$, and
	(\textbf{d}) $N_0$ on $\sqrt{s_{NN}}$. The squares, circles, triangles, and
	crosses represent the results from the $\pi^{\pm}$, $K^{\pm}$,
	$p(\bar p)$, and $\phi$ spectra, respectively. The closed symbols
	indicate the positive particles, and the open ones indicate
	negative particles.\label{f5}}
\end{figure*}

We call $T$ the effective temperature due to the fact that it
contains the contributions of thermal motion described by the
kinetic freeze-out temperature, $T_0$, and flow effect described
by the average transverse flow velocity, $\langle\beta_T\rangle$.
To relate the obtained $T$ to the expected $T_0$ and $\beta_T$,
one may consider a possible relation, $T_0=\langle
p_T\rangle/3.07$~\cite{50}, where $\langle p_T\rangle$ is the
average $p_T$, which can be obtained from the statistics in which
$T$, $q$, and $a_0$ play important roles. Then, we have
$\langle\beta_T\rangle=(2.07/3.07)\langle
p_T\rangle/\overline{m}$, where $\overline{m}$ is the average
energy (average moving mass) of the considered particle in the
source rest frame. It is regretful that we have no idea to
establish the linkage between $T$ ($T_0$) and the phase transition
temperature at present. In addition, $T$ from different
distributions (or functions or models) are different in their
sizes, though the trends are similar or
compatible~\cite{22a,22b,22c}.

There is a possible situation on the relation between $T_0$ and
$\langle\beta_T\rangle$, which satisfies the hydrodynamics in
which $T_0$ increases and $\langle\beta_T\rangle$ decreases with
the increase of particle mass. This is due to the early leaving
over for the massive particles during the evolution of the system.
The mass dependent $T$ reflects mainly the influence of flow
effect which shows a decreasing $\langle\beta_T\rangle$ with the
increase of mass. Although the contributions of $T_0$ and
$\langle\beta_T\rangle$ in $T$ are not dissociated in the present
work, we may obtain a definite $T$, which is not different from
the indefinite $T_0$ and $\langle\beta_T\rangle$ due to different
dissociated methods being used.

The boundary (5 GeV) from the quick to slow increases of $T$
reflects the change of reaction mechanism and/or generated matter.
There are two possible situations: (i) The products of the system
experience a change from baryon-dominated to meson-dominated,
where the hadron phase or nuclear matter always exists; (ii) The
system undergoes the de-confined phase transition from nuclear
matter to QGP. Indeed, a few GeV energy range is very important
due to it containing abundant information. This energy range
covers the initial energy of limiting fragmentation of nuclei, the
critical energy of possible change from baryon-dominated to
meson-dominated, and the critical energy of a possible de-confined
phase transition. Anyhow, the boundary should be given more
attention in the future study.

The entropy index $q$ reflects the degree of equilibrium or
non-equilibrium of the collision system. The system reaches the
equilibrium state at $q=1$, while $q \gg 1$ (e.g., $q>1.25$)
represents a non-equilibrium state. In our work, $q$ is close to
1, which shows that the equilibrium is basically maintained.
Usually, the equilibrium is relative. For an approximate equilibrium
situation, we can also use the concept of local equilibrium for
different local parts. If $q$ is not too large, for example, $q
\le 1.25$, the collision system is in approximate equilibrium or
local equilibrium~\cite{2,22}.

As can be seen from Figure 5, $q$ values are the highest for pions.
However, the pion production is the highest in AA collisions and one
would expect that they would reach equilibrium faster, which should
result in low $q$. However, in the collisions, the excitation of
pions is also the highest due to their small mass. This means that
pions are possibly further away from the equilibrium of original
partons than other particles, which results in high values of $q$
for~pions.

Because of most protons coming directly from the participant
nuclei, they have enough time to reach equilibrium during their
evolution. This also renders that $q$ is closer to 1 for emission
of protons. Our results show that $q$ increases with an increase in
the energy. At lower energy, the system is closer to the
equilibrium state because the lower energy evolution process is
slower, and the system has more time to reach equilibrium. From
initial collision to kinetic freeze-out, the evolution time is
very short. We have consistent results: the lower the collision
energy, the longer the evolution time, the closer to 1 the $q$,
and the more equilibrium the system has. Our results show indeed a $q$
closer to 1 at lower energy.

The parameter $a_0$ reflects the shape of a particle spectrum in
a low-$p_T$ region. If $a_0=1$ corresponds to a normal shape of the
spectrum, $a_0<1$ means a rising tendency and $a_0>1$ means a
falling tendency of the spectrum. Due to the constraint of
normalization, $a_0$ also affects the tendency of the spectrum in
intermediate- and high-$p_T$ regions, though it determines mainly
the tendency in low-$p_T$ region. The introduction of $a_0$
results in the fit process being more flexible, though one more
parameter is introduced. Although $a_0$ is dimensionless, its
introduction causes the dimension of $m_{T_i}$ to that of
$m_{T_i}^{a_0}$. This can be adjusted by the normalization
constant $C$ so that the probability density function is still
workable. Meanwhile, in Equations (1)--(3), the power $-q/(q-1)$
determines the thermodynamic consistency. The inconsistency or
approximate consistency caused by $a_0$ may be also adjusted by
the normalization constant.

The fact that $a_0\neq1$ means that the introduction of $a_0$ is
necessary. Our results show that $a_0$ is less than and close to 0
in most cases for the production of $\pi^{\pm}$, around 1 for the
production of $K^{\pm}$, around 1.5 for the production of $p(\bar
p)$, and around 1.5--3 for the production of $\phi$. Although the
meaning of difference in $a_0$ for different particles is not very
clear for us, $a_0<1$ for the production of $\pi^{\pm}$ renders
the contribution of resonance as significant, $a_0\approx1$
for the production of $K^{\pm}$ renders that the contribution of
resonance is not too large, and their production is not
restrained, while $a_0>1$ for the production of $p(\bar p)$ and
$\phi$ means that their productions are restrained. From low
energy to high ones, $a_0$ has slight fluctuations for given
particles in most cases. This is a reflection of the same or
similar shape of the spectrum in a low-$p_T$ region for given
particles in the considered energy~range.

Generally, with increasing $\sqrt{s_{NN}}$, $N_0$ increases
quickly and then slowly for produced particles which does not
include $p$. It is understandable that more energies are deposited
in the collisions at higher energy. Then, more particles are
produced due to the fact that the deposited energies are
transformed to masses due to the conservation of energy. The
situation of $p$ is different. As a component of projectile and
target nuclei, $p$ will be lost due to the collisions. The higher
the energy is, the more it is lost. The loss of $p$ will cause the
increase of other baryons due to the conversation of baryon
number. The increasing $N_0$ for produced particles also reflects
the increasing volume of the system.

For the most abundant produced particles, $\pi^{\pm}$ yield
increases quickly and then slowly. The boundary is around 5 GeV.
With increasing the mass, the boundary increases. This depends on
the threshold energy required for particle generation. Generally,
the average parameter is obtained by weighting the yields of
different particles. Because of the most abundant produced
particles being $\pi^{\pm}$ in collisions at high energy, the
average parameter is approximately determined by that for
$\pi^{\pm}$. Considering the massive yield of $p$ at low energy,
the average parameter is approximately determined by those for
$\pi^{\pm}$ and $p$.

To obtain an average parameter more accurately, one may consider
$\pi^{\pm}$, $K^{\pm}$, $p(\bar p)$, and other particles together.
The average parameters can be approximately used to fit the
spectra of different particles. In this case, the productions of
different particles are regarded as the result of simultaneous
decay of the system. Obviously, the application of average
parameters covers the mass dependent scenario, which reflects the
fine structure of the system evolution. In our opinion, the decay
of the system is not simultaneously. The massive particles are
produced early because they are left over in the hydrodynamics
of the system evolution.

Before summary and conclusions, we would like to point out that the
chemical potential mentioned in Equations (1)--(4) can be neglected at
high energy such as dozens of GeV and above. That is to say that
the chemical potential is redundant at high energy~\cite{51}.
However, the chemical potential is sizeable at low energy such as
a few GeV, though its influence is still small. At low energy, the
temperature values from the spectra of different particles overlap
each other, as what we observed in Figure \ref{f5}(a). The situations of
temperature values are nearly the same if we consider the two
cases of $\mu_{u,d}=0$ and $\mu_{u,d}=\mu_B/3$~\cite{4}. The
nearly independent of chemical potential renders that it is also
redundant at low energy~\cite{51}.

In addition, the influence of $q$ on the spectra in high-$p_T$
region at both the low and high energies is also remarkable. This
is not surprising that a slight increase of $q$ value can result
in a large increase in the high-$p_T$ region, while the values of
other parameters may remain nearly invariant. Because the value of
$q$ is close to 1, we may still say that the system stays in
approximate or local equilibrium, though a large difference of the
spectra in high-$p_T$ region is observed between the equilibrium
and approximate or local equilibrium. In our opinion, the
approximate or local equilibrium is achieved in the considered
collisions.
\\

{\section{Summary and Conclusion}}

We summarize here our main observations and conclusions.

(a)  We have used a new method to analyze the $p_T$ spectra of
identified particles produced in central AA collisions. The
particle's $p_T$ is regarded as the joint contribution of two
participant partons which obey the modified Tsallis-like
transverse momentum distribution and have random azimuths in
superposition. The Monte Carlo method is performed to calculate
and fit the experimental $p_T$ spectra of $\pi^{\pm}$, $K^{\pm}$,
$p(\bar p)$, and $\phi$ produced in central Au-Au and Pb-Pb
collisions over an energy range from 2.16 to 2760 GeV measured by
international collaborations. Three free parameters, the effective
temperature $T$, entropy index $q$, and revised index $a_0$ are
obtained.

(b) Our results show that, with the increase of $\sqrt{s_{NN}}$,
$T$ increases quickly and then slowly in the results from
$\pi^{\pm}$ and $K^{\pm}$ spectra. The boundary is around 5 GeV.
This energy is possibly the critical energy of a possible
de-confined phase transition from hadron matter to QGP. The values
of $q$ are close to 1 and have a slight increase with increasing
$\sqrt{s_{NN}}$. This result shows that the system is in
approximate equilibrium in the considered energy range and closer
to the equilibrium at lower energy. Generally, the values of $a_0$
are mass dependent and not energy dependent. The resonance
generation of $\pi^{\pm}$ and the constraints of other particles
in a low-$p_T$ region are reflected by the values of $a_0$.
\\
\\
{\bf Data Availability}

The data used to support the findings of this study are included
within the article and are cited at relevant places within the
text as references.
\\
\\
{\bf Ethical Approval}

The authors declare that they are in compliance with ethical
standards regarding the content of this paper.
\\
\\
{\bf Disclosure}

The funding agencies have no role in the design of the study; in
the collection, analysis, or interpretation of the data; in the
writing of the manuscript; or in the decision to publish the
results.
\\
\\
{\bf Conflicts of Interest}

The authors declare that there are no conflicts of interest
regarding the publication of this paper.
\\
\\
{\bf Acknowledgments}

The work was supported by the Shanxi Agricultural
University Ph.D. Research Startup Project under Grant No.
2021BQ103, and by the Fund for Shanxi ``1331 Project" Key Subjects
Construction.
\\

\end{document}